\UseRawInputEncoding 

\documentclass[aps,showpacs,onecolumn,notitlepage,floatfix,amsmath,amssymb,nofootinbib]{revtex4-1}

\usepackage{amssymb}
\usepackage{amsmath}
\usepackage{epsfig}
\usepackage{graphicx}
\usepackage{subfig}
\usepackage{color}
\usepackage{latexsym}
\usepackage{bm,latexsym}
\usepackage{mathrsfs}
\usepackage{float}
\usepackage{pbox}

\usepackage[usenames,dvipsnames]{xcolor}
\definecolor{orange}{cmyk}{0,0.5,1,0}

\begin{document}

\title{Gravitational analog of the canonical acoustic black hole in Einstein-scalar-Gauss-Bonnet theory}

\author{ Pedro Ca\~nate$^{1,}$$^{2}$ }
\email{pcannate@gmail.com, pedro.canate@unimeta.edu.co}

\author{Joseph Sultana$^{3}$}
\email{joseph.sultana@um.edu.mt}

\author{Demosthenes Kazanas$^{4}$}
\email{demos.kazanas-1@nasa.gov}

\affiliation{ $^{1}$Departamento de F\'isica  Te\'orica, Instituto de F\'isica, Universidade do Estado do Rio de Janeiro, Rua S\~ao Francisco Xavier 524, Maracan\~a,
CEP 20550-013, Rio de Janeiro, Brazil.\\
$^{2}$Escuela de Ingenier\'ia, Departamento de Ciencias B\'asicas, Corporaci\'on Universitaria del Meta, A.A. 3244, Villavicencio/Meta, Colombia.\\
$^{3}$Department of Mathematics, Faculty of Science, University of Malta, Msida, Malta.\\
$^{4}$Astrophysics Science Division, NASA/Goddard Space Flight Center, Greenbelt, MD 20771, USA.}

\begin{abstract}
In this work, in the context of modified gravity, a curved spacetime analogous to the ``canonical acoustic black hole" is constructed. The source is a self-interacting scalar field which is non-minimally coupled to gravity through the Gauss-Bonnet invariant. The scalar-Gauss-Bonnet coupling function is characterized by three positive parameters: $\sigma$ with units of $(length)$, $\mu$ with units of $(length)^{4}$, and a dimensionless parameter $s$, thus defining a three-parameter model for which the line element of canonical acoustic black hole is a solution.
The spacetime is equipped with spherical and static symmetry and has a single horizon determined in Schwarzschild coordinates by the region $r=\mu^{1/4}$. The solution admits a photon sphere at $r=(3\mu)^{1/4}$, and it is shown that in the
region $(3\mu)^{1/4}\leq r<\infty$ the scalar field satisfies the null, weak, and strong energy conditions.
Nonetheless, the model with $s=1$ has major physical relevance since for this case the scalar field is well defined in the entire region $r\geq\mu^{1/4}$,
while for $s\neq1$ the scalar field blows up on the horizon.
\end{abstract}



\maketitle

\section{Introduction}

Astrophysical black holes are one of the most interesting compact objects predicted by general relativity theory (GR), and also
have been extensively discussed in several alternative gravity theories (e.g., Dilaton gravity \cite{Dilaton}, Scalar-Tensor-Vector gravity \cite{STV}, $f(G)$-gravity \cite{fG}, $f(T)$-gravity \cite{fT}, $f(R)$-gravity \cite{fR}). The existence of astrophysical black holes has been confirmed from the recently detected gravitational waves by the LIGO and Virgo collaborations \cite{L&V} and with the first direct image of
photons around the supermassive black hole at the center of M87 by the Event Horizon Telescope collaboration \cite{EHT}.
On the other hand, Hawking \cite{Hawking} showed that black holes emit thermal radiation corresponding to a certain temperature (Hawking temperature). This is considered to be one of the most important aspects of black hole physics. Nevertheless, the Hawking temperature of astrophysical black holes ($\approx 10^{-8}\times \frac{M_{\odot}}{M}$ $^{\circ}\mathbf{K}$,
where $M_{\odot}$ is the solar mass, and $M$ is the mass of the BH) is much smaller than the temperature of the cosmic microwave background
($\approx 2.725$ $^{\circ}\mathbf{K}$). Consequently, it is difficult to observe the Hawking radiation emitted by astrophysical black holes.
However, the thermal emission is not only characteristic of astrophysical black holes; it is also characteristic of the black hole analogues \cite{Unruh,Unruh_Son,Unruh2,Unruh3}.
Because of this, one of the biggest motivations for studying black hole analogues is the possibility of an experimental verification of the analogue
Hawking radiation (see \cite{Unruh,Unruh_Son,Unruh2,Unruh3,BHanalog} for details).

In his landmark work, Unruh \cite{Unruh} started from the isentropic equation for smooth fluid flow $\vec{v}$ in a medium which has an equation of state
$p=p(\rho)$, and assuming that the flow is irrotational $\vec{v} = \nabla\phi$, he showed that the equation of motion for the small perturbations to fluid flow
$\delta\vec{v} = \nabla\Phi$, with $\Phi=\delta\phi$, is exactly the equation of motion for a scalar field $\nabla_{\nu}\nabla^{\nu}\Phi=0$.  In this case
$\nabla_{\nu}$ corresponds to the following metric,
\begin{equation}\label{Unruh}
ds^{2} =  \frac{ \rho }{ c_{_{s}}\!(\rho) }\left\{ - \left[ c^{2}_{_{s}}(\rho) - v^{2} \right]d\tau^{2} {\bf-} 2\vec{v}\!\cdot\!d\vec{x}d\tau +  d\vec{x}\!\cdot\!d\vec{x} \right\},
\end{equation}
where $c_{_{s}}\!(\rho)$ is the local velocity of sound, defined by $c^{2}_{_{s}} = \frac{\partial p}{\partial \rho }$, whereas $\tau$ is the laboratory time coordinate, and $\vec{x}=(x^{1},x^{2},x^{3})$ stands for the spatial coordinates. The above metric has an
effective Lorentzian geometry with metric components $g_{\mu\nu}$ functions of the local properties of the fluid flow.
Assuming $c_{_{s}}=constant$ (taking $c_{_{s}}=1$ without loss of generality), and that the background flow is a spherically symmetric, stationary, and
convergent flow, one gets the following line element,
\begin{equation}\label{SSUnruh}
ds^{2} =   \rho \left[ - \left( 1 - v^{2} \right)d\tilde{\tau}^{2} +  \frac{ d\chi^{2} }{ 1 - v^{2} }
+ \chi^{2}(d\theta^{2}  + \sin^{2}\theta d\varphi^{2}) \right].
\end{equation}
with $\tilde{\tau} = \tau + \int \frac{v}{1-v^{2}}d\chi$, where $\tilde{\tau}$ is like the Schwarzschild time coordinate, $\chi$ is like the Schwarzschild areal-radial coordinate, and $d\Omega^{2}_{(2)} = d\theta^{2}  + \sin^{2}\theta d\varphi^{2}$ is the line element on the unit 2-sphere. The class of solutions determined by (\ref{SSUnruh}) is considered as an acoustic black hole, where the acoustic horizon is located at $\chi=\chi_{_{h}}$ such that
$v(\chi_{_{h}})=1$.
An acoustic black hole forms when the velocity of the fluid exceeds the velocity of sound on some closed surface. That surface forms a sonic horizon, an exact sonic analog of a black hole horizon where the sound modes, or phonons, (rather than light waves) cannot escape the event horizon (see \cite{Unruh} for details).
Therefore, according to (\ref{SSUnruh}), the acoustic black hole region is determined by $v^2(\chi) > 1$, while the acoustic black hole exterior by $v^2(\chi)<1$.
By using the same arguments as for astrophysical black holes, it is found that
the acoustic black hole (\ref{SSUnruh}) produces thermal radiation with a temperature given by,
$T = \frac{1}{2\pi}\frac{dv}{d\chi} \big|_{v=1}$; for a review, see \cite{Unruh,Unruh_Son}.

{\it Canonical acoustic black hole (CABH):} a remarkable particular case of (\ref{SSUnruh}), for a spherically symmetric flow of incompressible fluid, called canonical acoustic black hole, was found by Visser \cite{Visser}.
Since the fluid is incompressible, it follows that the background density is position independent $\rho=\rho_{_{0}}$, which, together with the barotropic assumption $p = p(\rho)$
yields that the pressure is also position independent $p=p_{_{0}}$. 
Subsequently, the continuity equation
$\partial_{\tau}\rho + \vec{\nabla}\cdot(\rho\vec{v})=0$, implies $v^{2} =\mathcal{C}/\chi^{4}$ where $\mathcal{C}$ is a integration constant. 
The Visser solution for the canonical acoustic black hole is obtained by substituting $\rho=\rho_{_{0}}$, and $v^{2}=\mathcal{C}/\chi^{4}$, in the line element (\ref{SSUnruh}), 
\begin{equation}
ds^{2} = - \left( 1 - \frac{\mathcal{C}}{\chi^{4}} \right)\left(\rho_{_{0}}^{1/2}d\tilde{\tau}\right)^{2} + \left( 1 - \frac{\mathcal{C}}{\chi^{4}} \right)^{\!-1}\left(\rho_{_{0}}^{1/2}d\chi\right)^{2}
+ \left(\rho_{_{0}}^{1/2}\chi\right)^{2}(d\theta^{2}  + \sin^{2}\theta d\varphi^{2}).
\end{equation}
One can define a normalization constant $\mu = \mathcal{C}\rho_{_{0}}^{2}$, and the re-scaling coordinates $t = \rho_{_{0}}^{1/2}\tilde{\tau}$, $r=\rho_{_{0}}^{1/2} \chi$. Therefore, the Visser solution is given by, 
\begin{equation}\label{NewBH}
ds^{2} = - \left( 1 - \frac{\mu}{r^{4}} \right)dt^{2} + \left( 1 - \frac{\mu}{r^{4}} \right)^{\!-1}dr^{2}
+ r^{2}(d\theta^{2}  + \sin^{2}\theta d\varphi^{2}).
\end{equation}
This line element describes a $(3+1)-$dimensional acoustic black hole with acoustic event horizon determined by the region $r_h=\mu^{1/4}$. This solution has been extensively studied, and its properties like quasinormal modes \cite{Qn_modes}, scattering of sound \cite{Scatt_CABH}, and analogue Hawking radiation \cite{H_Rad_CABH}, have all been thoroughly investigated.
However, the line element (\ref{NewBH}) is distinct from any of the black hole geometries typically considered in GR.
In fact, at the end of the next section, we will show that in the framework of GR, in order to generate a  spacetime with metric given by (\ref{NewBH}), one
needs to use a matter field source for which all known energy conditions are violated at every point in spacetime.

The outline of the paper is as follows: In the next section the field equations for the Einstein-scalar-Gauss-Bonnet (EsGB) gravity are derived, and also the line element of CABH as an exact solution of GR is discussed.  In Sec. \ref{CABH_EsGB} we show that the line element for the CABH is a solution of pure EsGB, and some features of the solution in the EsGB frame are discussed.
In this paper we use units where $G = k_{B} = c = \hbar = 1$, and the metric signature ($-+++$) is used throughout.
\section{ Einstein-scalar-Gauss-Bonnet gravity }\label{F_EQS}
The Einstein-scalar-Gauss-Bonnet theory (EsGB), also called extended scalar-tensor-Gauss-Bonnet theory \cite{scalariz_Doneva}, is defined by the following action,
\begin{equation}\label{actionL}
S[g_{ab},\phi,\psi_{a}] = \int d^{4}x \sqrt{-g} \left\{ \frac{1}{16\pi}\left(R - \frac{1}{2}\partial_{\mu}\phi\partial^{\mu}\phi  - 2 \mathscr{U}(\phi) + \boldsymbol{f}(\phi) R_{_{GB}}^{2} \right) + \frac{1}{4\pi}\mathcal{L}_{\rm matter}(g_{ab},\psi_{a})  \right\},
\end{equation}
where the term $R$ corresponds to the Ricci scalar  and by itself defines the Einstein-Hilbert Lagrangian density; the term $2\mathcal{L}_{_{\rm SF}}(g_{ab},\phi) =  \frac{1}{2}\partial_{\mu}\phi\partial^{\mu}\phi + 2 \mathscr{U}(\phi)$ corresponds to the Lagrangian density for the self-interacting scalar field; whereas
$\boldsymbol{f}(\phi)R_{_{GB}}^{2}$ stands for the Gauss-Bonnet invariant coupled to the scalar field,
with $\boldsymbol{f}(\phi)$ being the coupling function, and $R_{_{GB}}^{2}$ is a higher curvature term called Gauss-Bonnet invariant (or quadratic Gauss-Bonnet term) and is defined in terms of the curvature invariants as
$R_{_{GB}}^{2} = R_{\alpha\beta\mu\nu}R^{\alpha\beta\mu\nu} - 4R_{\alpha\beta}R^{\alpha\beta} + R^{2}$.
The Lagrangian density $\mathcal{L}_{\rm matter}(g_{ab},\psi_{a})$ represents the rest of the matter fields that can appear in the theory.
The EsGB field equations that are derived from the theory (\ref{actionL}), are the following:
\begin{equation}\label{modifEqs}
G_{\alpha}{}^{\beta} + \Theta_{\alpha}{}^{\beta} =  8\pi (E_{\alpha}{}^{\beta} + T_{\alpha}{}^{\beta}), 
\end{equation}
where $G_{\alpha}{}^{\beta} = R_{\alpha}{}^{\beta} - \frac{R}{2} \delta_{\alpha}{}^{\beta}$ are the components of the Einstein tensor, whereas $\Theta_{\alpha}{}^{\beta}$ and $E_{\alpha}{}^{\beta}$ are defined by,
\begin{equation}\label{E_GB}
\Theta_{\alpha}{}^{\beta} =  \frac{1}{2}( g_{\alpha\rho} \delta_{\lambda}{}^{\beta} + g_{\alpha\lambda} \delta_{\rho}{}^{\beta})\eta^{\mu\lambda\nu\sigma}\tilde{R}^{\rho\xi}{}_{\nu\sigma}\nabla_{\xi}\partial_{\mu}\boldsymbol{f}(\phi),\quad\quad
  8\pi E_{\alpha}{}^{\beta} = -\frac{1}{4}(\partial_{\mu}\phi\partial^{\mu}\phi)\delta_{\alpha}{}^{\beta} + \frac{1}{2}\partial_{\alpha}\phi \partial^{\beta}\phi - \mathscr{U}(\phi)\delta_{\alpha}{}^{\beta},
\end{equation}
with $\tilde{R}^{\rho\gamma}{}_{\mu\nu} = \eta^{\rho\gamma\sigma\tau}R_{\sigma\tau\mu\nu} = \epsilon^{\rho\gamma\sigma\tau}R_{\sigma\tau\mu\nu}/\sqrt{-g}$. Thus, the quantities  $\Theta_{\alpha}{}^{\beta}$ are the components of a tensor which we refer to as the Gauss-Bonnet curvature tensor, since that captures the contribution to the spacetime curvature due to the effects of the GB term; $E_{\alpha}{}^{\beta}$ are the components of the canonical energy-momentum tensor of self-interacting scalar field, while $T_{\alpha}{}^{\beta}$ are the components of the energy-momentum tensor associated with $\mathcal{L}_{\rm matter}(g_{ab},\psi_{a})$.
Finally, the scalar-field equation of motion and the conservation law for the matter fields $\psi_{a}$, are given by,
\begin{equation}\label{scalar_Eq}
\nabla^{2}\phi + \dot{\boldsymbol{f}}(\phi) R_{_{GB}}^{2} = 2 \dot{\mathscr{U}}(\phi), \quad\quad\quad\quad  \nabla_{\alpha}T^{\alpha\beta} = 0,
\end{equation}
where the dot over $\boldsymbol{f}$ and $\mathscr{U}$ denotes their derivatives with respect to the scalar
field, i.e., ($\dot{\boldsymbol{f}} = \frac{d\boldsymbol{f}}{d\phi}$, $\dot{\mathscr{U}} = \frac{d\mathscr{U}}{d\phi}$).
Thus, the Equations (\ref{modifEqs}) and (\ref{scalar_Eq}) are the basic equations for EsGB gravity with
additional matter fields.\\
Various cosmological applications of the EsGB theory have been studied in detail recently. For instance,
it was shown in \cite{cosmic_acceleration} that the theory can describe the present stage of cosmic acceleration, and can lead to an exit from a scaling matter-dominated epoch to a late-time accelerated expansion \cite{cosmos}. The possible reconstruction of the coupling and potential functions for a given scale factor was considered in \cite{reconstruction}, and the consequences of EsGB in an inflationary setting have been considered in \cite{inflaton}.
Exact black hole solutions of EsGB theory are scarce.
Instead, numerical black hole solutions have been constructed \cite{scalariz_Doneva,Evasion}.
\subsection{EsGB field equations for the static and spherically symmetric spacetime}
Staticity and  spherical symmetry (SSS) are one of the simplest symmetries
with which a spacetime is usually equipped in order to obtain analytic, or numerical black hole solutions. For an asymptotically flat, static and spherically symmetric black hole (AF-SSS-BH) solution
of the set of Eqs. (\ref{modifEqs}) and (\ref{scalar_Eq}), with non-trivial functions $\phi$ and $\boldsymbol{f}(\phi)$, we can assume that the scalar field is static and spherically symmetric, $\phi = \phi(r)$, and also that the metric takes the static and spherically symmetric form,
\begin{equation}\label{SSSmet}
ds^{2} =  - e^{ A(r) }dt^{2} + e^{ B(r) }dr^{2}  + r^{2}(d\theta^{2}  + \sin^{2}\theta d\varphi^{2}),
\end{equation}
with $A = A(r)$ and $B = B(r)$ being unknown functions depending only on $r$.
Below, we include the explicit form of the field equations assuming both the SSS ans\"atz for the metric
(\ref{SSSmet}) and SSS scalar field $\phi(r)$.

The non-vanishing Einstein tensor components are given by,
\begin{equation}\label{GabSSS}
G_{t}{}^{t}\!=\!\frac{ e^{^{\!\!-B}} }{ r^{2} }\!\!\left( -rB' - e^{^{\!B}} + 1 \right)\!, \hspace{0.25cm}  G_{r}{}^{r} \!=\! \frac{ e^{^{\!\! -B}} }{ r^{2} }\!\!\left( rA' - e^{^{\!B}} + 1 \right)\!, \hspace{0.25cm} G_{\theta}{}^{\theta}\!=\! G_{\varphi}{}^{\varphi}\!=\!\frac{ e^{^{\!\!-B}} }{ 4r }\!\!\left( rA'^{2} - rA'B' + 2rA'' + 2A' - 2B' \right)\!,
\end{equation}
where the prime denotes the derivative with respect to the radial coordinate $r$. The non-vanishing components of the GB curvature tensor with arbitrary coupling function $\boldsymbol{f}(\phi)$ are,
\begin{eqnarray}
&& \Theta_{t}{}^{t} = \frac{ e^{ -2B} }{ 4r^{2} }\left\{ 16(e^{B} - 1)\ddot{\boldsymbol{f}}\phi'^{2} - 8[ (e^{B} - 3)B'\phi' - 2(e^{B} - 1)\phi'']\dot{\boldsymbol{f}} \right\}, \quad\quad \Theta_{r}{}^{r} =    \frac{2(e^{B} - 3)e^{-2B}A'\phi'\dot{\boldsymbol{f}} }{r^{2}}    \label{GBtt} \\
&& \Theta_{\theta}{}^{\theta} = \Theta_{\varphi}{}^{\varphi} = \frac{ e^{ -2B} }{ 4r } \left\{   - 8A'\ddot{\boldsymbol{f}}\phi'^{2} - 4\left[ (A'^{2} + 2A'')\phi' + (2\phi'' - 3B'\phi')A'\right]\dot{\boldsymbol{f}} \right\} \label{GBthth}.
\end{eqnarray}
The non-trivial components of the energy-momentum tensor of self-interacting scalar field are given by,
\begin{equation}\label{EttyErr}
8\pi E_{t}{}^{t} = 8\pi E_{\theta}{}^{\theta} = 8\pi E_{\varphi}{}^{\varphi}   = -\frac{1}{ 4 } e^{ -B} \phi'^{2}  - \mathscr{U}(\phi), \quad\quad\quad\quad 8\pi E_{r}{}^{r} = \frac{1}{ 4 } e^{ -B} \phi'^{2} - \mathscr{U}(\phi). \end{equation}
Through the components (\ref{GabSSS})-(\ref{EttyErr}) and the field equations (\ref{modifEqs}), it follows that the non-trivial $T_{\alpha}{}^{\beta}$-components are;
$T_{t}{}^{t}$, $T_{r}{}^{r}$, $T_{\theta}{}^{\theta}=T_{\varphi}{}^{\varphi}$.
Inserting the above given components in the field equations (\ref{modifEqs}) written as
$\mathbf{C}_{\alpha}{}^{\beta} = G_{\alpha}{}^{\beta} + \Theta_{\alpha}{}^{\beta} - 8\pi(E_{\alpha}{}^{\beta} + T_{\alpha}{}^{\beta}) = 0$, we obtain that the EsGB field equations for SSS spacetime, take the form:
\begin{eqnarray}
&&\!\mathbf{C}_{t}{}^{t}\!=\!0\hspace{0.01cm}\!\Rightarrow\!\hspace{0.01cm}
4e^{B}\!\!\left( rB' \!+\! e^{ B} \!-\! 1 \right)\!-\!\left[ r^{2}e^{B}\!+\!16(e^{ B}\!-\!1)\ddot{\boldsymbol{f}}\right]\!\phi'^{2}\!+\!8\!\left[ (e^{ B}\!-\!3)B'\phi'\!-\!2(e^{B}\!-\!1)\phi'' \right]\!\dot{\boldsymbol{f}}
\!-\!4r^{2}e^{2B}(\mathscr{U}\!-\!8\pi T_{t}{}^{t} )\!=\!0, \label{Eqt}\\
&&\nonumber\\
&&\!\mathbf{C}_{r}{}^{r}\!=\!0\hspace{0.01cm}\!\Rightarrow\!\hspace{0.01cm} 4e^{B}\left( -rA'\!+\!e^{ B}\!-\!1 \right)\!+\!r^{2}e^{B}\phi'^{2}\!-\!8(e^{ B}\!-\!3)A'\phi'\dot{\boldsymbol{f}}\!-\! 4r^{2}e^{2B}(\mathscr{U}\!-\! 8\pi T_{r}{}^{r})\!=\!0,
\label{Eqr}\\
&&\nonumber\\
&&\!\mathbf{C}_{\theta}{}^{\theta}\!=\!0\hspace{0.01cm}\!\Rightarrow\!\hspace{0.01cm} e^{B}\!\!\left\{ rA'^{2} \!-\!2B'\!+\!(2\!-\!rB')A'\!+\!2rA''\right\}\!+\!re^{B}\phi'^{2}\!-\!8A'\ddot{\boldsymbol{f}}\phi'^{2} \nonumber \\
&& \hskip7.16cm - 4\left[ (A'^{2}\!+\!2A'')\phi'\!+\!(2\phi''\!-\!3B'\phi')A' \right]\dot{\boldsymbol{f}} \!+\!4 r e^{2B}(\mathscr{U}\!-\!8\pi T_{\theta}{}^{\theta})\!=\!0.\label{Eqte}
\end{eqnarray}
whereas the equations (\ref{scalar_Eq}) becomes,
\begin{equation}\label{phi2}
2r\phi'' + (4 + rA' - rB')\phi' + \frac{4e^{-B}\dot{\boldsymbol{f}}}{r} \left[ (e^{B} - 3)A'B' - (e^{B} - 1)(2A'' + A'^{2})\right] - 4r e^{B}\dot{\mathscr{U}} = 0, \quad \nabla_{\alpha}T^{\alpha\beta} = 0.
\end{equation}

The problem of describing a AF-SSS-BH within the framework of EsGB gravity reduces to solving the field equations (\ref{Eqt}), (\ref{Eqr}), (\ref{Eqte}) and (\ref{phi2}) for the SSS metric ans\"atz (\ref{SSSmet}) with:
\begin{equation}
e^{A(r)} = \left( 1 - \frac{2\mathscr{M}\!(r)}{r}\right)e^{2\delta(r)}, \quad\quad\quad\quad e^{B(r)} = \left( 1 - \frac{2\mathscr{M}\!(r)}{r}\right)^{-1}.
\end{equation}
where $\mathscr{M}\!(r)$ and $\delta(r)$ are well-defined functions at the region $r\geq r_{h}$, where $r=r_{h}$ is the event horizon defined by $\mathscr{M}\!(r_{h}) = r_{h}/2$. Furthermore, in order to have an asymptotically flat spacetime, it is required that $\lim\limits_{r\rightarrow \infty}\mathscr{M}\!(r) = \mathscr{M}_{_{\infty}}$, and $\lim\limits_{r\rightarrow \infty}\delta(r) = \delta_{_{\infty}}$, where $\mathscr{M}_{_{\infty}}$ and $\delta_{_{\infty}}$ real constants\footnote{The static symmetry of metric ans\"atz (\ref{SSSmet}) implies that without loss of generality the constant $\delta_{_{\infty}}$
can be taken as $\delta_{_{\infty}}=0$.}.
According to (\ref{GBtt})-(\ref{GBthth}), the non-triviality of the functions
$\phi$ and $\boldsymbol{f}(\phi)$ is necessary. Otherwise, the contribution of GB-term vanishes, and the EsGB field equations are reduced to the Einstein-scalar field equations of general relativity.  Furthermore,
the scalar field must be a well defined function in the region $r\geq r_{h}$, whereas the  asymptotic flat behavior of the spacetime implies the following boundary condition
$\lim\limits_{r\rightarrow \infty}\phi(r) = \phi_{\!_{\infty}}$, where $\phi_{\!_{\infty}}$ is a real number that denotes the asymptotic value of the scalar field  (see \cite{Evasion} for details).
On the other hand, since the additional matter fields includes more complicated fields for which the field equations may be more complicated to solve, in this work we limit ourselves to the case of (pure) EsGB gravity\footnote{In \cite{Sols_sour}, exact solutions of EsGB with electromagnetic sources are presented.}, i.e., the field equations (\ref{Eqt})-(\ref{phi2}) with $T_{\alpha}{}^{\beta}=0$.
\subsection{Analogue of CABH in GR: Violation of energy conditions at every point in spacetime}
In GR without cosmological constant, $G_{\alpha}{}^{\beta} = 8\pi T_{\alpha}{}^{\beta}$, the only spherically symmetric solution of the vacuum ($T_{\alpha}{}^{\beta}=0$) field equations is the Schwarzschild solution. This follows from Birkhoff's theorem which states that the static, asymptotically flat solution is described only in terms of one single conserved quantity measured at infinity, namely the ADM mass \cite{ADMmass}. Hence, in order for the line element (\ref{NewBH}) to be a solution of the Einstein field equations, one needs to consider the contribution of a matter field coupled to the theory. \\
Below we show that for the line element (\ref{NewBH}), the energy-momentum tensor associated with this matter field violates the weak energy condition (WEC)
everywhere in the spacetime.
The WEC states that for any timelike vector $\boldsymbol{k} = k^{\mu}\partial_{\mu}$, (i.e., $k_{\mu}k^{\mu}<0$), the energy-momentum tensor $T_{\mu\nu}$ obeys the inequality
$T_{\mu\nu}k^{\mu}k^{\nu} \geq 0$, which means that the local energy density $\rho_{\!_{_{loc}}}= T_{\mu\nu}k^{\mu}k^{\nu}$ as measured by any observer with timelike vector $\boldsymbol{k}$ is a non-negative quantity. Following \cite{WEC}, for a diagonal energy-momentum tensor $(T_{\alpha\beta})=diag \left( T_{tt},T_{rr},T_{\theta\theta},T_{\varphi\varphi} \right)$, which can be
conveniently written as,
\begin{equation}\label{diagonalEab}
T_{\alpha}{}^{\beta} = - \rho \hskip.05cm \delta_{\alpha}{}^{t}\delta_{t}{}^{\beta} + P_{r} \hskip.05cm \delta_{\alpha}{}^{r}\delta_{r}{}^{\beta} + P_{\theta} \hskip.05cm \delta_{\alpha}{}^{\theta}\delta_{\theta}{}^{\beta} + P_{\varphi} \hskip.05cm \delta_{\alpha}{}^{\varphi}\delta_{\varphi}{}^{\beta} ,
\end{equation}
where $P_i$ are the pressures along the $r$, $\theta$ and $\varphi$ directions,
the WEC leads to,
\begin{equation}\label{WEC}
\rho = - T_{t}{}^{t} \geq0, \quad \rho + P_{a} \geq0, \quad  a = \{ r, \theta, \varphi \}.
\end{equation}
According to GR theory the matter distribution, in the spacetime equipped with the metric (\ref{NewBH}),
is related to the metric by the Einstein field equations $T_{\alpha}{}^{\beta}=G_{\alpha}{}^{\beta}/(8\pi)$. Thence, for the line element (\ref{NewBH}) in the GR context, the non-zero $T_{\alpha}{}^{\beta}$ components are; $T_{t}{}^{t} = T_{r}{}^{r} = \frac{3\mu}{8\pi r^{6}}$, and  $T_{\theta}^{\theta} = T_{\varphi}^{\varphi} = -\frac{6\mu}{8\pi r^{6}}$.
Therefore, by identification with (\ref{diagonalEab}), it follows that $\rho + P_{r} =0$ together with,
\begin{equation}\label{V_WEC}
\rho = - T_{t}{}^{t} = -\frac{3\mu}{8\pi r^{6}}, \quad\quad\quad \rho + P_{\theta} = \rho + P_{\varphi} = -\frac{9\mu}{8\pi r^{6}}.
\end{equation}
The expressions (\ref{V_WEC}) are negative for any value of the radial coordinate $r$, which implies that the weak energy condition (\ref{WEC}) and the null energy condition\footnote{The NEC states that for any null vector, $n^{\alpha}$, $T_{\mu\nu}n^{\mu}n^{\nu}\geq0$. In terms of (\ref{diagonalEab}) the NEC implies: $\rho + P_{a} \geq0$,  $a = \{ r, \theta, \varphi \}$} (NEC), are violated everywhere in
 the spacetime.
In a similar way one can show that the strong energy
condition\footnote{The SEC states that for any timelike vector $k^{\alpha}$, $(T_{\mu\nu} - \frac{1}{2}T_{\alpha}{}^{\alpha}g_{\mu\nu} )k^{\mu}k^{\nu}\geq0$. In terms of (\ref{diagonalEab}) the SEC implies: $\rho + P_{a} \geq0$ and  $\rho + \sum_{a} P_{a} \geq0$, $a = \{ r, \theta, \varphi \}$. } (SEC), and dominant energy condition\footnote{The DEC states that for any timelike vector $n^{\alpha}$, $T_{\mu\nu}k^{\mu}k^{\nu}\geq0$, and  $T_{\mu\nu}k^{\mu}$ is not spacelike. In terms of (\ref{diagonalEab}) the DEC implies: $\rho\geq0$ and $P_{a}\in[-\rho,\rho]$ for all $a = \{ r, \theta, \varphi \}$.} (DEC) are also violated for any value
of the radial coordinate.
\section{Gravitational analogue of CABH in (pure) EsGB theory}\label{CABH_EsGB}
The particular scalar-Gauss-Bonnet coupling function in terms of the radial coordinate, which defines the three-parameter EsGB-model used to derive the gravitational analog of CABH, is given by the following function:
\begin{equation}\label{fNew}
\boldsymbol{f}(\phi(r)) \!=\! - \frac{1}{12\mu}\left[  \frac{r^{6}}{2} -  \int^{r}_{ \sigma }   \frac{ \chi^{8} }{ ( \chi^{4} - \mu)^{^{\!\!\frac{5}{6}}} }\left( \int^{\chi}_{s\mu^{\frac{1}{4}}}  \frac{ d\xi }{ ( \xi^{4} - \mu)^{^{\!\!\frac{1}{6}}} } \right) d\chi \right],
\end{equation}
where $\sigma$, $\mu$, and $s$ are positive real parameters of the model; $\sigma$ with units of $(length)$, $\mu$ with units of $(length)^{4}$; while $s$ is dimensionless.
For this coupling function the (pure) EsGB field equations, (\ref{Eqt})-(\ref{phi2}) with $T_{\alpha}{}^{\beta}=0$, are fulfilled by the line element (\ref{NewBH}), with scalar field $\phi$ and potential $\mathscr{U}$, given by:
\begin{eqnarray}
 && \phi(r) = \int^{r}_{\sigma} \sqrt{ \frac{ 2(14\chi^{4} - 15\mu) }{ 3\chi^{2}( \chi^{4}-\mu ) } -  \frac{ 4\chi(7\chi^{4} - 12\mu) }{ 9( \chi^{4}-\mu )^{^{\!\!\frac{11}{6}}} }\left( \int^{\chi}_{s\mu^{\frac{1}{4}}}   \frac{ d\xi }{ ( \xi^{4} - \mu)^{^{\!\!\frac{1}{6}}} } \right)  } \hspace{0.1cm}d\chi, \label{phiNew}\\
&&\mathscr{U}(\phi(r)) = \frac{ 3\mu - 10r^{4} }{ 6r^{6}}  + \frac{ 5r^{4} - 6\mu }{ 9r^{3}( r^{4}-\mu )^{^{\!\!\frac{5}{6}}} } \int^{r}_{s\mu^{\frac{1}{4}}} \frac{ d\xi }{ ( \xi^{4} - \mu)^{^{\!\!\frac{1}{6}}} }. \label{UNew}
\end{eqnarray}
Thus, the line element (\ref{NewBH}) can be  reinterpreted
as a solution of a pure EsGB model with, according to (\ref{actionL}), a Lagrangian density determined by (\ref{fNew}) and (\ref{UNew}). 
For this solution the curvature invariants are,
\begin{equation}\label{invariants}
R = \frac{6\mu}{r^{6}}, \quad\quad\quad R_{\alpha\beta}R^{\alpha\beta} = \frac{90\mu^{2}}{r^{12}}, \quad\quad\quad R_{\alpha\beta\sigma\nu}R^{\alpha\beta\sigma\nu} = \frac{468\mu^{2}}{r^{12}},
\end{equation}
showing that the spacetime metric  is well-behaved in the region $r\in(0,\infty)$. Moreover,
as in the spherically symmetric (vacuum or electrovacuum) GR solutions, the origin $r=0$ is a physical singularity where the curvature invariants (\ref{invariants})  diverge.
For the case $\mu>0$ the solution possesses a single horizon ($r_{h}$) determined by the region $r=r_{h}=\mu^{\frac{1}{4}}$. The metric (\ref{NewBH}) has $\delta(r)=0$, and $\mathscr{M}\!(r) = \mu/r^{3}$, and then $\lim\limits_{r\rightarrow \infty}\mathscr{M}\!(r) = 0$, indicating  that spacetime is AF.\\
On the other hand, according to the curvature invariants (\ref{invariants}), this vacuum EsGB black hole solution yields $R_{_{GB}}^{2} = 144 \mu^{2}/r^{12}$. Whereas,
 the non-vanishing components of the Gauss-Bonnet curvature tensor, (\ref{GBtt}) and (\ref{GBthth}), are given by 
\begin{eqnarray}
&&\Theta_{t}{}^{t}\!=\!\frac{1}{9r^{6}}\!\!\left[ - 3(2r^{4} + 3\mu) + \frac{ 2r^{3}(r^{4}-3\mu)}{(r^{4}-\mu)^{\frac{5}{6}}}\!\!\int^{r}_{s \mu^{\frac{1}{4}}}\!\!\frac{ d\xi }{ ( \xi^{4} - \mu)^{^{\!\!\frac{1}{6}}} }  \right]\!,\quad \Theta_{r}{}^{r}\!=\! \frac{2(2r^{4}-3\mu)}{3r^{6}}\!\!\left[ 3 - \frac{r^{3}}{(r^{4} - \mu)^{\frac{5}{6}}}\!\!\int^{r}_{s\mu^{\frac{1}{4}}}\!\!\frac{ d\xi }{ ( \xi^{4} - \mu)^{^{\!\!\frac{1}{6}}} } \right]\!, \label{sGBt_t} \\
&&\Theta_{\theta}{}^{\theta} \!=\! \Theta_{\varphi}{}^{\varphi} \!=\!\frac{2}{9r^{6}}\!\!\left[ - 3(r^{4}-12\mu) + \frac{ r^{3}(r^{4}-3\mu)}{(r^{4}-\mu)^{\frac{5}{6}}}\!\!\int^{r}_{s\mu^{\frac{1}{4}}}\!\!\frac{ d\xi }{ ( \xi^{4} - \mu)^{^{\!\!\frac{1}{6}}} } \right]\!\label{sGBth_th}.
\end{eqnarray}
{\bf Trivial Case:} when $\mu=0$ the scalar field (\ref{phiNew}) and the potential (\ref{UNew}) vanish, $\phi(r)=0=\mathscr{U}(r)$, the metric (\ref{NewBH}) becomes the Minkowski metric, while the coupling function (\ref{fNew}) diverges. Despite the divergence of $\dot{\boldsymbol{f}}$, when $\mu =0$ the components of the Gauss-Bonnet curvature tensor, (\ref{sGBt_t}) and (\ref{sGBth_th}), vanish i.e.
$\Theta_{\alpha}{}^{\beta} = 0$.
On the other hand, the quantities $\dot{\boldsymbol{f}}R_{_{GB}}^{2}$ and $\dot{\mathscr{U}}$, which are involved in the equation of motion of the scalar field
(\ref{scalar_Eq}), satisfy $\lim\limits_{\mu\rightarrow0}\dot{\mathscr{U}}(r)=0$, and $\lim\limits_{\mu\rightarrow0}\dot{\boldsymbol{f}}R_{_{GB}}^{2}= 0$  
since $R_{_{GB}}^{2} = 144 \mu^{2}/r^{12}$.
Therefore, for the EsGB solution (\ref{NewBH}), (\ref{fNew}), (\ref{phiNew}), and (\ref{UNew}), with $\mu=0$, the contribution to the spacetime curvature due to the effects of the sGB term is null. 

\begin{figure}
\centering
\epsfig{file=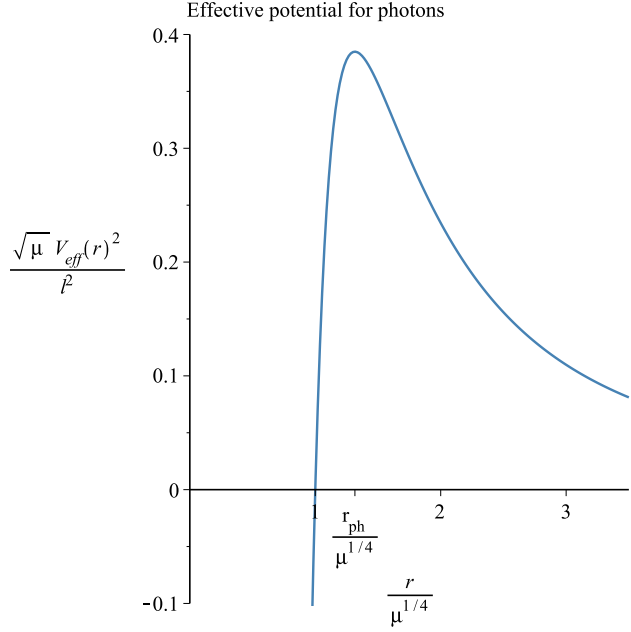, scale=0.44}
\caption{ \label{V_eff} Effective potential for a massless test particle, in the gravitational analog of CABH geometry, is illustrated. The illustration is for any positive value of $\mu$, $\mu>0$; the abscissa is, $r/\mu^{\frac{1}{4}}$; and the ordinate is, $\frac{\mu^{\frac{1}{2}}}{\ell^{2}}\mathsf{V}_{\!e\!f\!f}(r)$. The effective potential blow up at $r=0$, is zero at $r=r_{h}=\mu^{\frac{1}{4}}$, and becomes maximum at $r= r_{\!\!_{ph}}=(3\mu)^{\frac{1}{4}}$.
}
\end{figure}

\begin{figure}
 \centering
{
    \includegraphics[width=0.45\textwidth]{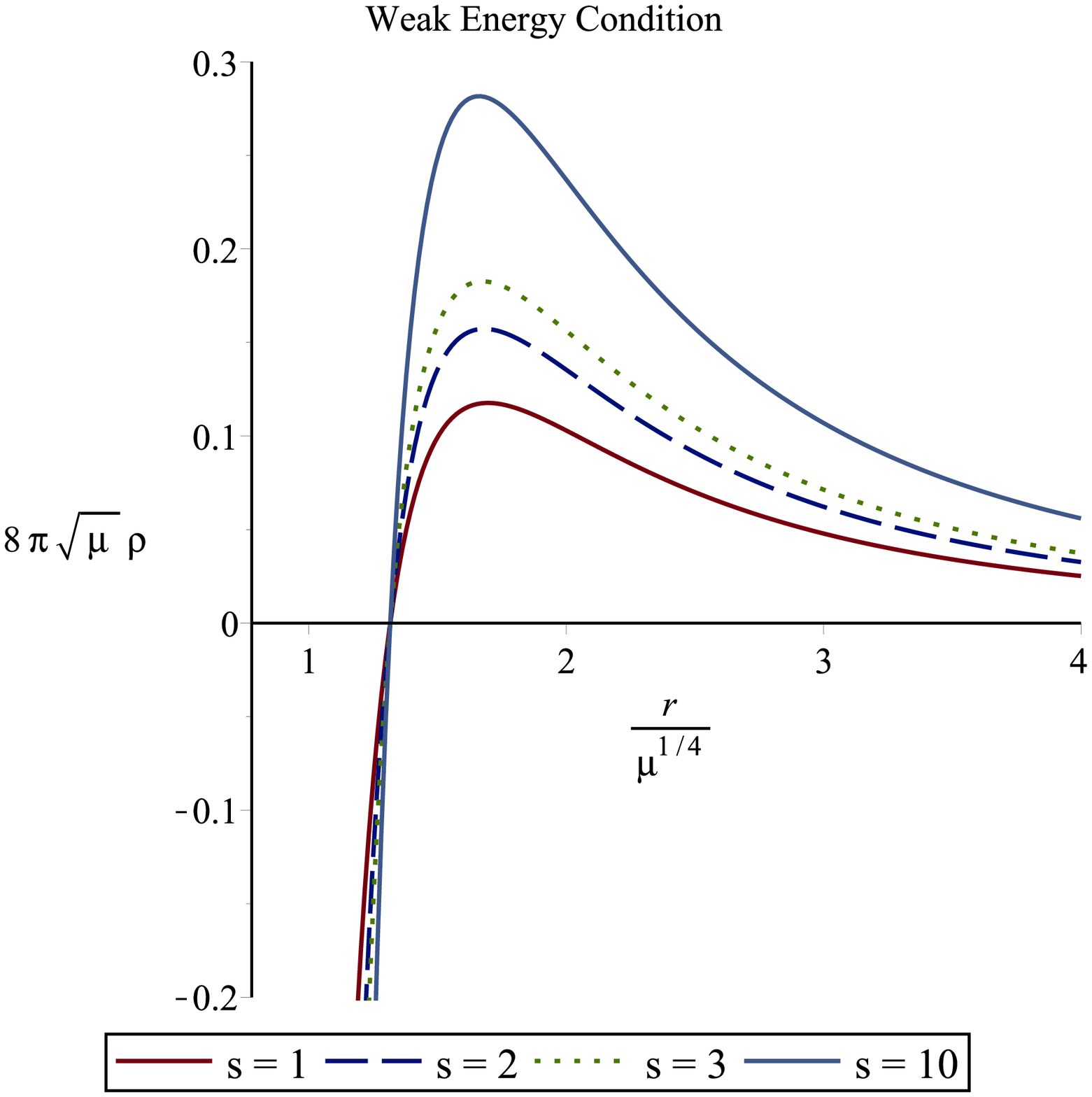}}
{
    \includegraphics[width=0.45\textwidth]{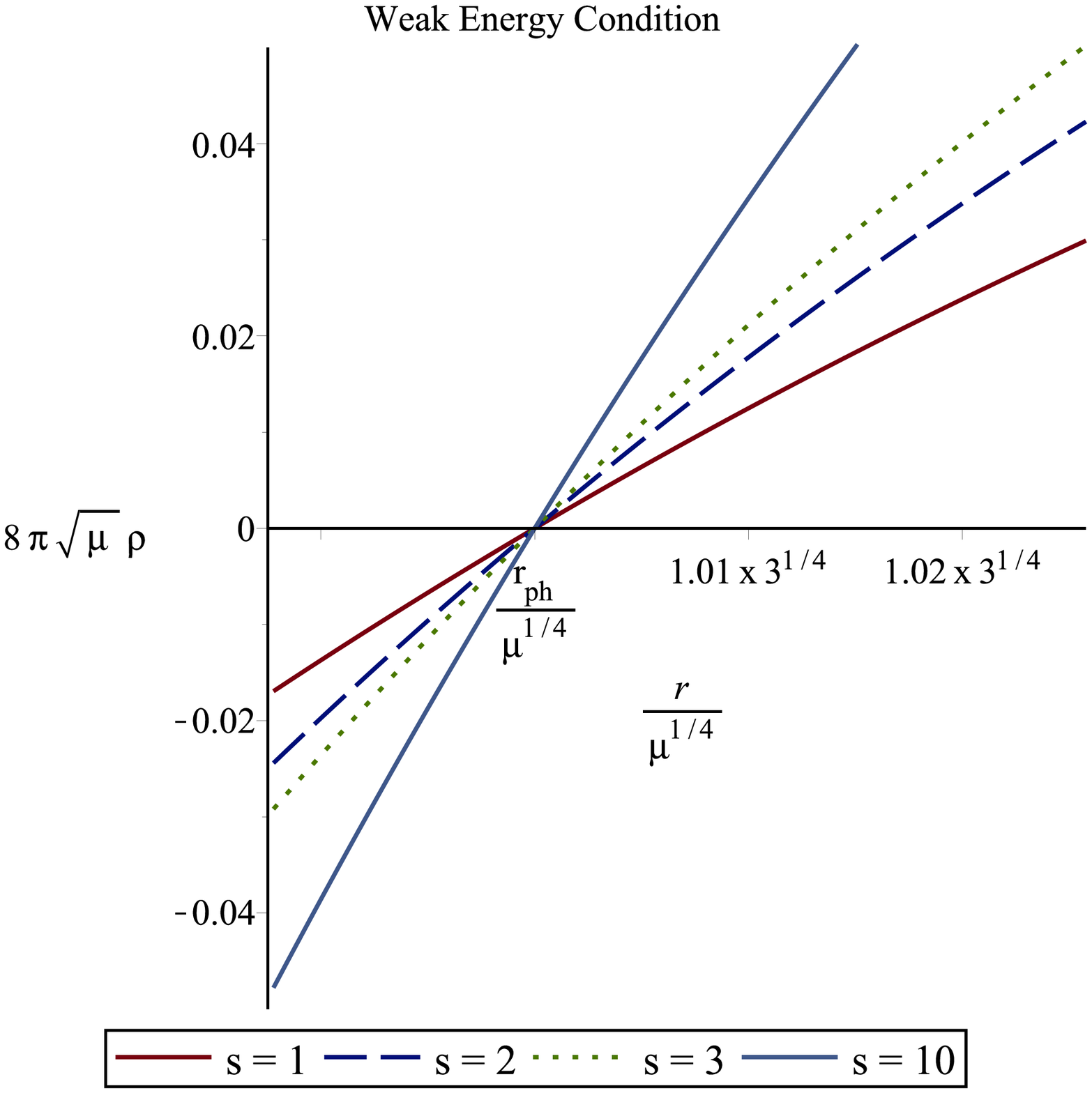}}
\caption{ \label{EnergyrhoZ} Behavior of $\mu^{\frac{1}{2}}\rho(\tilde{r})$ in the exterior region $\tilde{r}>1$, with different values of $s$. To the left, the range of the  horizontal axis is: $\tilde{r}\in[0.8,4]$. To
the right, the range of the  horizontal axis is:
$\tilde{r}\in[1.3,1.35]$, where this is a neighborhood around of $\tilde{r}_{\!\!_{ph}}=3^{\frac{1}{4}}\approx1.316074$. }
 \label{Energyrho}
\end{figure}

\begin{figure}
\centering
\epsfig{file=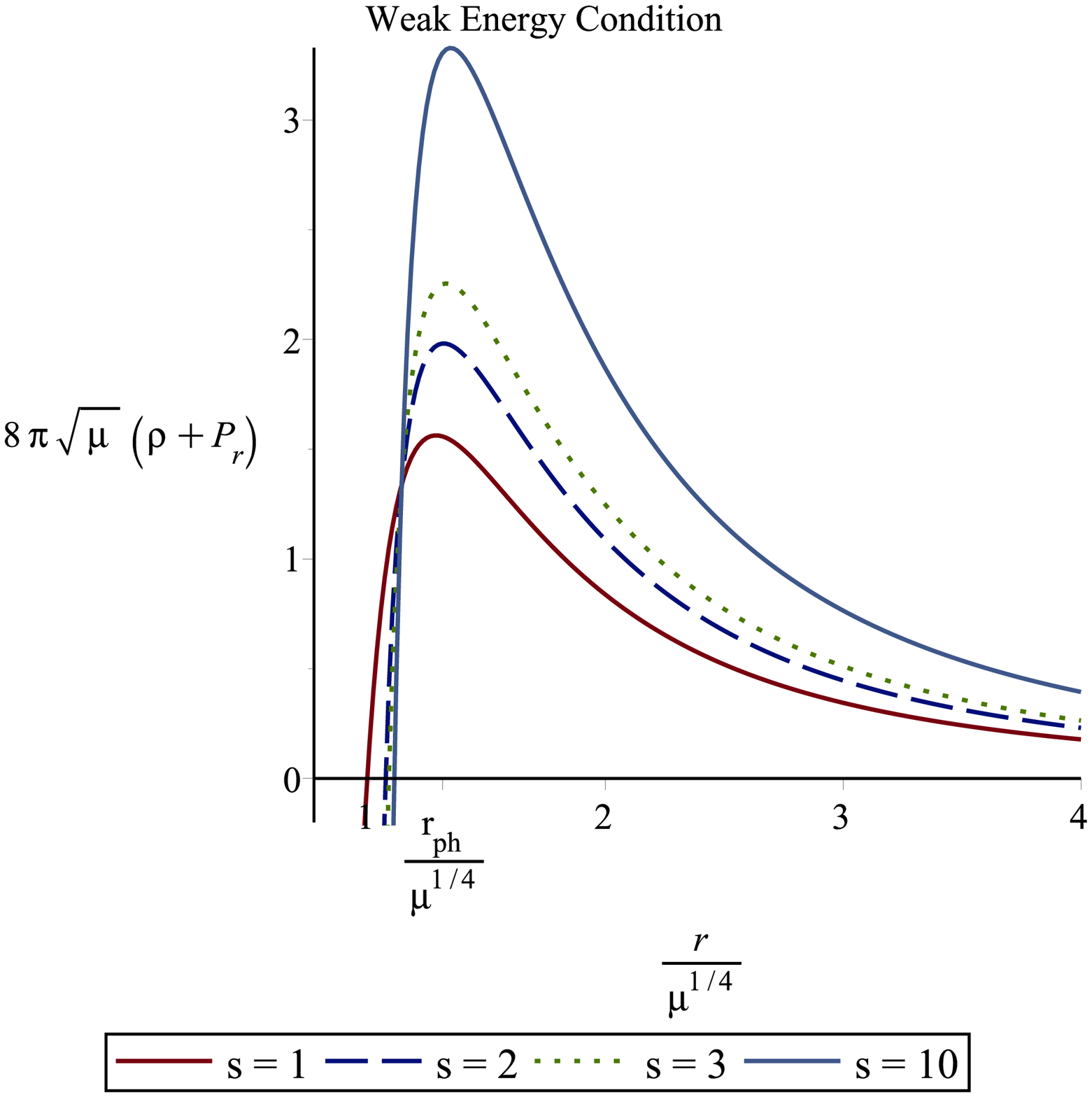, scale=0.5}
\caption{\label{EnergyrhoPr} Behavior of $8\pi\mu^{\frac{1}{2}}[\rho(\tilde{r})+P_{r}(\tilde{r})]$ in the exterior region $\tilde{r}>1$, with different values of $s$. The range of the  horizontal axis is: $\tilde{r}\in[0.8,4]$.
}
\end{figure}

\begin{figure}
\centering
\epsfig{file=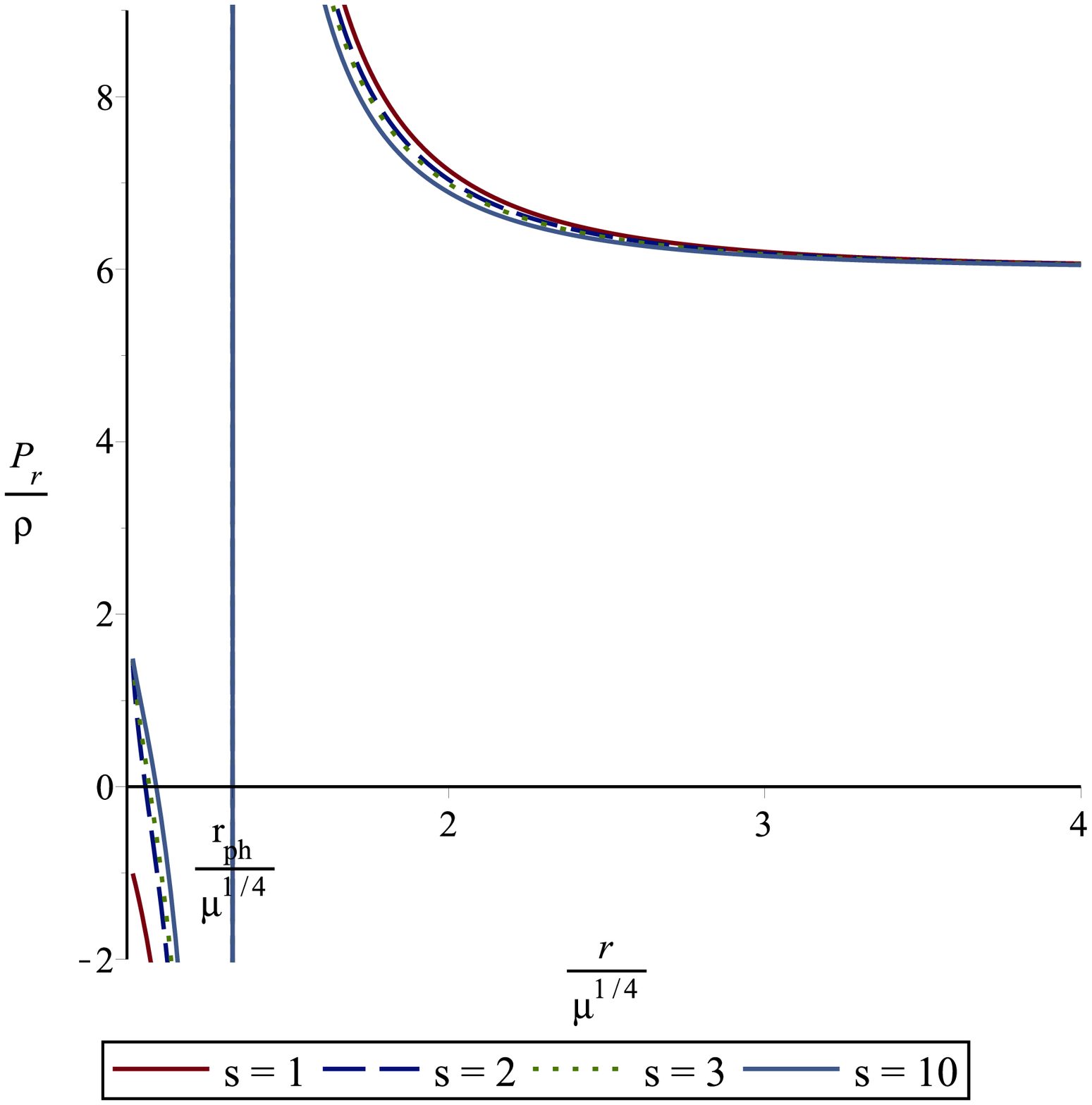, scale=0.5}
\caption{\label{Pr_p} Behavior of $P_{r}(\tilde{r})/\rho(\tilde{r})$ in the exterior region $\tilde{r}>1$, with different values of $s$. The range of the horizontal axis is: $\tilde{r}\in[1,4]$. The vertical line $\tilde{r}=\tilde{r}_{_{ph}}/\mu^{\frac{1}{4}}=3^{\frac{1}{4}}$  correspond to the localization of the photon sphere.
}
\end{figure}

\begin{figure}
\centering
\epsfig{file=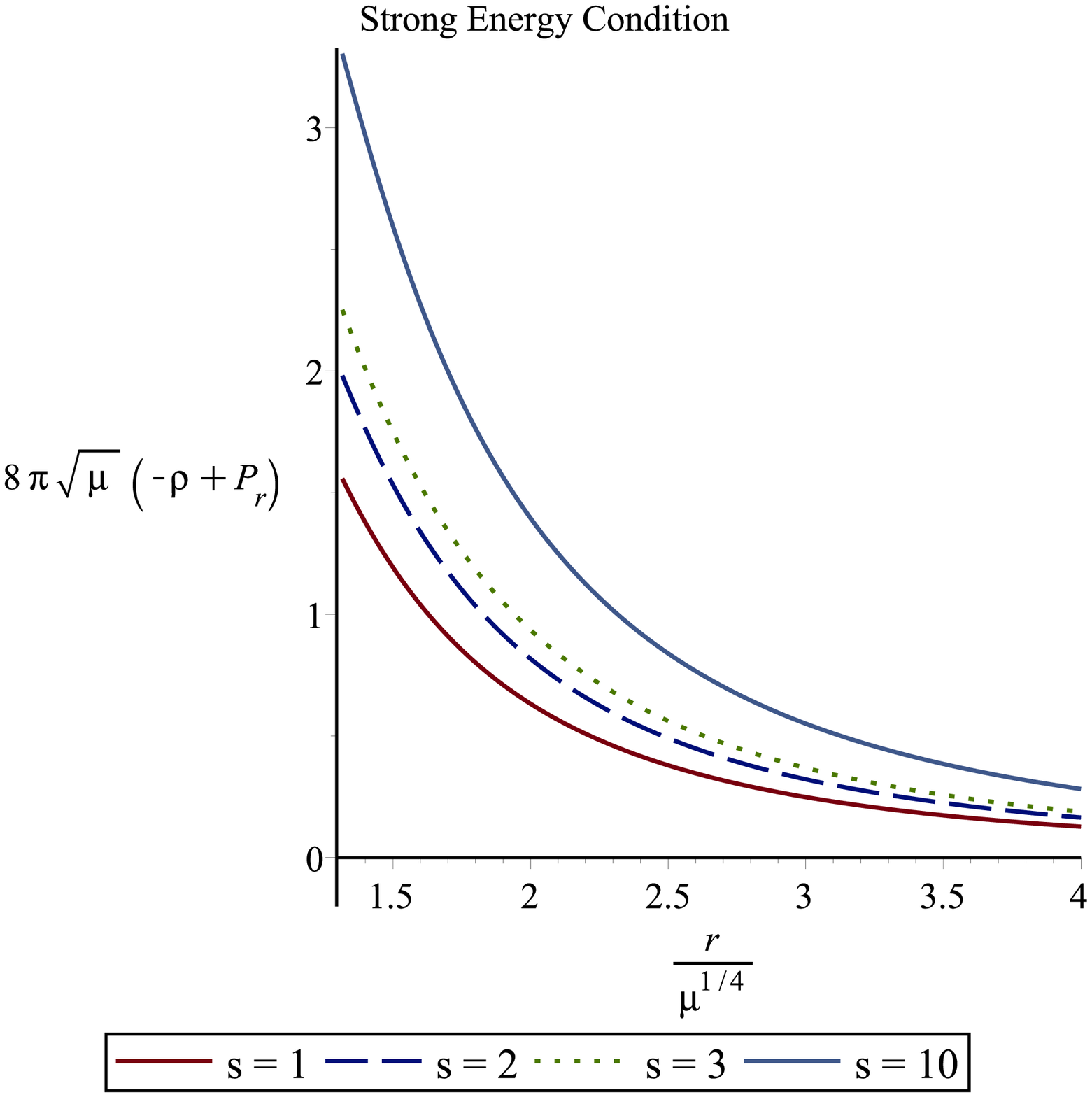, scale=0.5}
\caption{\label{SEC} Behavior of $8\pi\mu^{\frac{1}{2}}[-\rho(\tilde{r})+P_{r}(\tilde{r})]$ in the exterior region $\tilde{r}>3^{1/4}$, with different values of $s$. The range of the horizontal axis is: $\tilde{r}\in[3^{\frac{1}{4}},4].$
}
\end{figure}

\begin{figure}
\centering
\epsfig{file=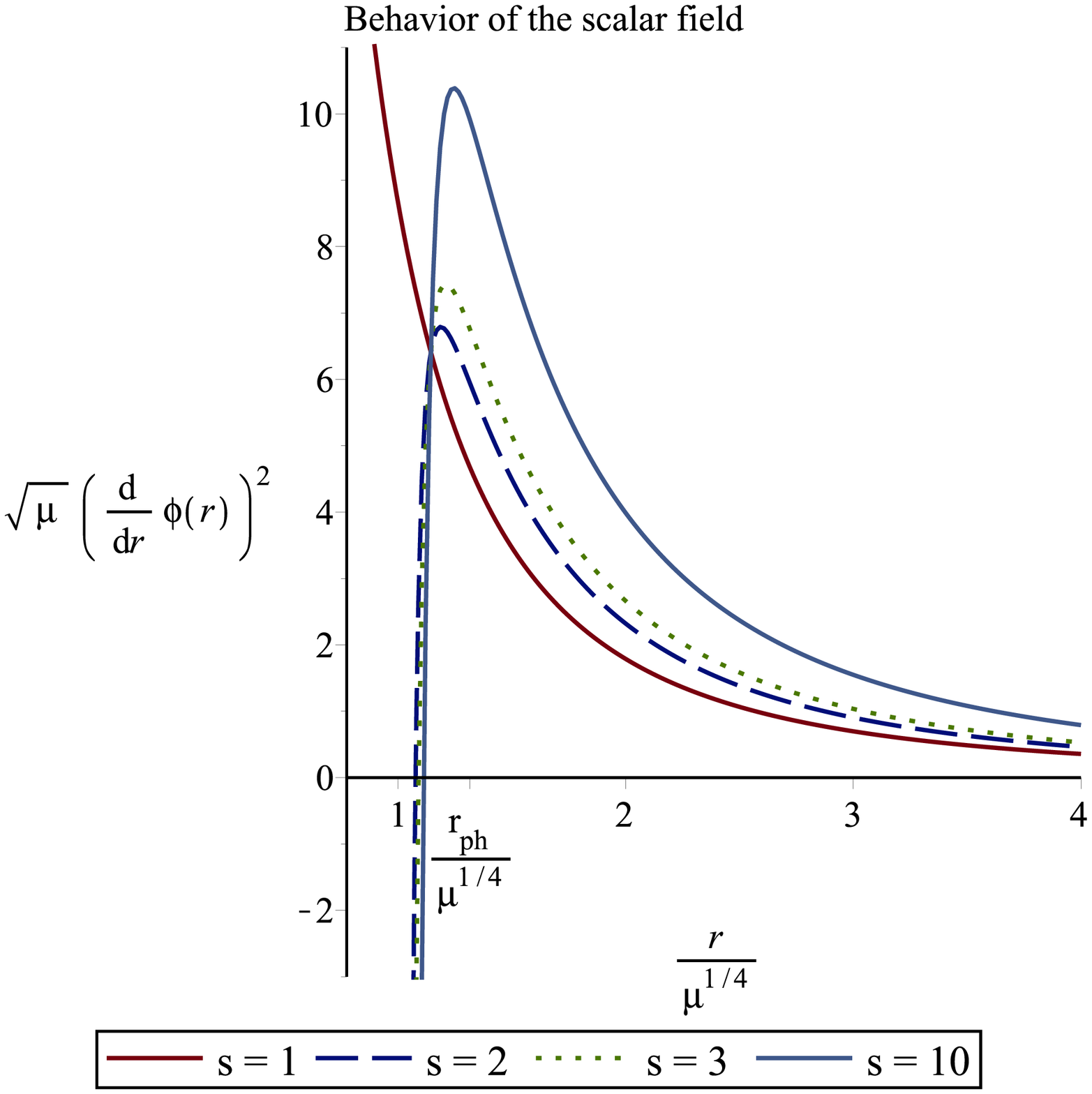, scale=0.5}
\caption{\label{D1phi} Behavior of  $\mu^{1/2}\left(\frac{d\phi}{dr}\right)^{2}$ in the exterior region $\tilde{r}>1$, for different values of $s$. The case $s=1$ is the most relevant since $\mu^{1/2}\left(\frac{d\phi}{dr}\right)^{2}$ is well defined in all domain of the radial coordinate. Whereas for the cases $s\neq1$, the quantities $\mu^{1/2}\left(\frac{d\phi}{dr}\right)^{2}$ blow up in the horizon $\tilde{r}=1$. The range of the horizontal axis is: $\tilde{r}\in[0.8,4]$.
}
\end{figure}

{\bf Null geodesics and capture cross-section for light:} for the spacetime metric (\ref{NewBH}), we will analyze the behavior of light rays in the region $r>r_{h}$.
\\
For null geodesics, the photon trajectories are described by the equations:
\begin{equation}\label{nullgeo}
\frac{d^{2}x^{\alpha}}{d\lambda^{2}} + \Gamma^{\alpha}_{\beta\nu} \frac{dx^{\beta}}{d\lambda}\frac{dx^{\nu}}{d\lambda} = 0, \quad\quad \textup{ and } \quad\quad ds^{2}(u,u) = 0, \end{equation}
where $\lambda$ and $u = \frac{ dx^{\alpha} }{ d\lambda }\partial_{ \alpha } $ are the affine parameter and the
null vector to the  null geodesics, respectively. Symmetries imply the conservation of the energy $\mathsf{E} = \left( 1 - \frac{\mu}{r^{4}} \right)\frac{dt}{d\lambda}$,
and angular momenta $\ell  =  r^{2}\frac{d\varphi}{d\lambda}$ and $\mathsf{L}_{\theta} =  r^{2}\frac{d\theta}{d\lambda}$. 
The conservation of the angular momentum $\ell$, or of $\mathsf{L}_{\theta}$, means that the particle will move on a plane. Thus, without loss of generality one can choose the plane $\theta=\pi/2$, and then introducing the constants of motion, $\mathsf{E}$ and $\ell$, in the Eqs. (\ref{nullgeo}), yields,
\begin{equation}
- \left( 1 - \frac{\mu}{r^{4}} \right)\!\!\left(\frac{dt}{d\lambda}\right)^{2} + \left( 1 - \frac{\mu}{r^{4}} \right)^{\!-1}\!\!\left(\frac{dr}{d\lambda}\right)^{2}
+ r^{2}\!\!\left(\frac{d\varphi}{d\lambda}\right)^{2} = 0 \quad \Rightarrow \quad \left(\frac{dr}{d\lambda}\right)^{2} + \frac{\ell^{2}}{r^{2}}\!\!\left( 1 - \frac{\mu}{r^{4}} \right) = \mathsf{E}^{2}
\end{equation}
which can be written as $\left( \frac{dr}{d\lambda} \right)^{2} + \mathsf{V}_{\!e\!f\!f}^{2}(r) = \mathsf{E}^{2}$, with the effective potential given   by
$\mathsf{V}_{\!e\!f\!f}^{2}(r) = \frac{\ell^{2}}{r^{2}}\!\!\left( 1 - \frac{\mu}{r^{4}} \right)$.
The effective potential is maximum at $r= r_{\!\!_{ph}} = (3\mu)^{\frac{1}{4}}$, $\mathsf{V}_{\!e\!f\!f}(r_{\!\!_{ph}})=\mathsf{V}_{\!e\!f\!f}^{Max}=\frac{2^{\frac{1}{2}}\ell}{3^{\frac{1}{2}}(3\mu)^{\frac{1}{4}}}$. Thus,  $\mathsf{E}=\mathsf{V}_{\!e\!f\!f}(r_{\!\!_{ph}})$ corresponds to an unstable circular orbit of radius $r_{\!\!_{ph}} =3^{\frac{1}{4}}r_{h}=(3\mu)^{\frac{1}{4}}$. This orbit is called the
photon circle (or last photon orbit) (for details see\cite{pho_orb}). This is an interesting feature of black holes. By definition, the last photon orbit is a
region of black hole spacetime where the spacetime curvature around them is so strong that photons are forced to travel in closed orbits. Due to the spherical
symmetry, the condition $\mathsf{E}=\mathsf{V}_{\!e\!f\!f}(r_{\!\!_{ph}})$ defines a collection of infinitely many such orbits, therefore the last photon orbit
is also called photon sphere \cite{pho_orb}.
On the other hand, an incoming photon with energy $\mathsf{E} > \mathsf{V}_{\!e\!f\!f}^{Max}$,
enters $r=\mu^{\frac{1}{4}}$, so it is captured by the hole. While an incoming photon with $\mathsf{E} < \mathsf{V}_{\!e\!f\!f}^{Max}$, is scattered by the potential to infinity, so it is not captured.
According to \cite{Wald}, the impact parameter $\boldsymbol{b}$  will be $\boldsymbol{b} = \ell/\mathsf{V}_{\!e\!f\!f}^{Max} = 3^{\frac{3}{4}}\mu^{\frac{1}{4}}/2^{\frac{1}{2}}$, and hence the
capture cross-section for a light beam is $\sigma=\pi\boldsymbol{b}^{2}=3^{\frac{3}{2}}\pi r_{h}^{2}/2$.
The plot of the effective potential illustrates the above-described behavior; see Fig. \ref{V_eff}.
The photon sphere can cast a black hole shadow for an observer at infinity. This is a disk specified by its radius $r_{sh}$ and it gives an apparent size and shape of the black hole
which has been observed recently for the first time \cite{EHT} for the supermassive black hole in the giant elliptical galaxy M87 using the Event Horizon
Telescope. For a static spherically symmetric asymptotically flat black hole $r_{sh}$ is just the impact parameter $\boldsymbol{b}$. So for example, for a
Schwarzschild black hole $r_{sh} = 3\sqrt{3}M = \frac{3\sqrt{3}}{2}r_h$, where $M$ is the geometric mass of the black hole. It has been shown \cite{hod17,lu19}
that for a static black hole in Einstein gravity $r_{sh} \leq 3\sqrt{3}M$. In other words the Schwarzschild black hole casts the largest shadow. In our case for
the CABH we get $r_{sh} = \frac{3^{3/4}}{\sqrt{2}}r_h$. Hence in terms of the event horizon $r_h$, which gives the actual size of any black hole, the shadow
of the CABH is smaller than that of the Schwarzschild black hole.
%
%
\\
{\bf Local energy density of the scalar field:} the non-zero components of the energy-momentum tensor of the scalar field system (\ref{phiNew})-(\ref{UNew}), are given by,
\begin{eqnarray}
&& 8\pi E_{t}{}^{t} = 8\pi E_{\theta}{}^{\theta} = 8\pi E_{\varphi}{}^{\varphi} = -\frac{2(r^{4}-3\mu)}{9r^{6}(r^{4} - \mu)^{\frac{5}{6}}}\left[ 3(r^{4} - \mu)^{\frac{5}{6}} - r^{3}\int^{r}_{s\mu^{\frac{1}{4}}} \frac{ d\xi }{ ( \xi^{4} - \mu)^{^{\!\!\frac{1}{6}}} }      \right]\\
&& 8\pi E_{r}{}^{r} = \frac{ 1  }{ r^{6} }\left[ 4r^{4} - 3\mu + \frac{2r^{3}(3\mu  - 2r^{4})}{3(r^{4} - \mu)^{\frac{5}{6}}}\int^{r}_{s\mu^{\frac{1}{4}}} \frac{ d\xi }{ ( \xi^{4} - \mu)^{^{\!\!\frac{1}{6}}} } \right].
\end{eqnarray}
Hence, by identification with (\ref{diagonalEab}) and defining $\tilde{r} = r/\mu^{\frac{1}{4}}$, is follows that,
\begin{eqnarray}
&& 8\pi \rho(\tilde{r}) = - 8\pi P_{\theta}(\tilde{r}) = - 8\pi P_{\varphi}(\tilde{r}) = \frac{2(\tilde{r}^{4}-3) }{ 9 \mu^{\frac{1}{2}}\tilde{r}^{6}(\tilde{r}^{4} - 1)^{\frac{5}{6}}}\left[ 3(\tilde{r}^{4} - 1)^{\frac{5}{6}} - \tilde{r}^{3}\int^{\tilde{r}}_{s} \frac{ d\xi }{ ( \xi^{4} - 1)^{^{\!\!\frac{1}{6}}} } \right],\label{rho_phi} \\
&& 8\pi P_{r}(\tilde{r}) = \frac{1}{ \mu^{\frac{1}{2}}\tilde{r}^{6} }\left[  4\tilde{r}^{4} - 3 + \frac{2\tilde{r}^{3}(3-2\tilde{r}^{4})}{3(\tilde{r}^{4} - 1)^{\frac{5}{6}}}\int^{\tilde{r}}_{s} \frac{ d\xi }{ ( \xi^{4} - 1)^{^{\!\!\frac{1}{6}}} }\right].\label{Pr_phi}
\end{eqnarray}
From (\ref{rho_phi}), for all $s$, it follows that; $\rho(\tilde{r}_{\!\!_{ph}}) = 0$; and
$\rho + P_{\theta} = \rho + P_{\varphi}=0$, which is consistent with (\ref{WEC}). The behaviors of quantities
$\rho =-E_{t}{}^{t}$ and $\rho + P_{r} = -E_{t}{}^{t}+E_{r}{}^{r}$ (for different values of $s$) in the exterior region $\tilde{r} = r/\mu^{\frac{1}{4}} > 1$, where $\tilde{r}=1$ is
the localization of the horizon, are shown in the Figs. \ref{Energyrho} and \ref{EnergyrhoPr} respectively.
%
%
It is found that these quantities are positive-definite in the region $3^{\frac{1}{4}} \leq \tilde{r} < \infty$, where $\tilde{r} = \tilde{r}_{\!\!_{ph}} = 3^{\frac{1}{4}}$ is the position of the photon sphere.
Additionally, according to the correspondence between the scalar field energy-momentum tensor given by (\ref{EttyErr}), with an effective fluid stress-energy tensor given by (\ref{diagonalEab}), one may define the effective equation of state $w_{r}=P_{r}(\tilde{r})/\rho(\tilde{r})$. The behavior of $w_{r}$ in the exterior region $\tilde{r}>1$ is presented in Fig. \ref{Pr_p}, indicating that the inequality $w_{r}>0$ is holding outside of the photon sphere.\\
For the SEC one gets $\rho + \sum_{a} P_{a} = -\rho +  P_{r}$, according to (\ref{rho_phi})-(\ref{Pr_phi}). The quantity $-\rho +  P_{r}$ is positive-definite in the region $3^{\frac{1}{4}}\leq \tilde{r} < \infty$, cf. Fig. (\ref{SEC}).
Hence in the region $r_{\!\!_{ph}} \leq r < \infty$ the canonical energy-momentum tensor of the self-interacting scalar field satisfies the NEC, WEC, and SEC.\\
On the other hand, in Fig. \ref{D1phi}, the behavior or the real-valued function $(\phi'(r))^{2}$ for several values of the parameter $s$, are illustrated. It is
found that for any value of $s$ the function $(\phi'(r))^{2}$ goes to zero as $r$ tends to infinity. This means that the scalar field (\ref{phiNew}), goes to a
constant in the asymtotic flat region, $\lim\limits_{r\rightarrow \infty}\phi(r) = \phi_{\!_{\infty}}$.
However, it is also shown that $s=1$ is the highlighted case, because it is the only one for which $(\phi'(r))^{2}$ is positive throughout the region $0<r<\infty$.  Now considering the case
$s=1$, we now turn to the form of the scalar field at near distances from the horizon. Expressing the integrand of (\ref{phiNew}) in terms of a power series, we get
\begin{equation}
\phi(r)\!=\!\int^{r}_{\sigma}\!\frac{3\sqrt{2915}}{ 55 \mu^{\frac{1}{4}} }\!\!\left[1\!-\!\frac{1002}{901}\!\frac{ (\chi\!-\!\mu^{\frac{1}{4}}) }{ \mu^{\frac{1}{4}} } \!+\! \frac{20998074}{18671423 }\frac{(\chi\!-\!\mu^{\frac{1}{4}})^{2}}{ \mu^{\frac{1}{2}} } \!-\! \frac{536957427763}{487865611567}\frac{(\chi\!-\!\mu^{\frac{1}{4}})^{3}}{\mu^{\frac{3}{4}}}
\!+\!\mathcal{O}\!\left(\!(\chi-\mu^{\frac{1}{4}})^{4}\!\right)\!\right]\!d\chi,
\end{equation}
such that the integration leads to
\begin{equation}\label{Newphi_h}
\phi(r)\!=\!\phi_{\!_{h}}+\frac{3\sqrt{2915}}{ 55 \mu^{\frac{1}{4}} }\!\!\left[\!(r\!-\!\mu^{\frac{1}{4}})\!-\! \frac{501}{901}\!\frac{(r\!-\!\mu^{\frac{1}{4}})^{2}}{ \mu^{\frac{1}{4}} } \!+\! \frac{6999358}{18671423}\!\frac{(r\!-\!\mu^{\frac{1}{4}})^{3}}{ \mu^{\frac{1}{2}} } \!-\! \frac{536957427763}{1951462446268}\!\frac{(r\!-\!\mu^{\frac{1}{4}})^{4}}{ \mu^{\frac{3}{4}} }\!\right] \!+ \mathcal{O}\!\left(\!(r-r_{h})^{5} \right)\!.
\end{equation}
The above expression is valid for $r\approx\mu^{\frac{1}{4}}$, where $\phi_{\!_{h}}$ a real constant determined by $\sigma$ and $\mu$, indicating that $\phi(r)$ is well defined at the horizon $\phi(r_{h})=\phi_{h}$.
%
%

\section{Conclusion}
In Einstein-scalar-Gauss-Bonnet theory, we have investigated the construction of a spacetime analog to the canonical acoustic black hole. The EsGB-model for which the line element of the CABH is a solution, is defined by a self-interacting scalar field (\ref{phiNew})-(\ref{UNew}), and a coupling function (\ref{fNew}) characterized by three positive-parameters: $\sigma$, $\mu$ and $s$.
The energy conditions are violated by the scalar field in a limited region of the spacetime, while outside of the photon sphere $r_{\!_{ph}}\leq r<\infty$ the NEC, WEC and SEC are fulfilled. In contrast in
GR this line element leads to an energy-momentum tensor for which all known energy conditions are violated at every point of the spacetime.
Moreover, another interesting feature that we have found is that in the case $s=1$, the scalar field $\phi(r)$, and its radial derivative $\phi'(r)$, are both well-defined on the entire region $r_{h}\leq r<\infty$. Whereas for $s\neq1$ the scalar field and its radial derivative, diverge on the horizon.

In view of the results obtained in this work, it would be interesting to establish the relationship between the properties of acoustic black holes with those of black holes in the framework of EsGB gravity. This will be explored in a future in a future paper.

\textbf{Acknowledgments}: P. C. is supported by Coordenaç\~ao de Aperfeiçoamento de Pessoal de N\'ivel Superior-Brasil (CAPES) - C\'odigo de Financiamento 001.

\section{Appendix: relevant functions }
In this appendix we include the explicit form of the components of the GB-Tensor asociated with the spacetime metric (\ref{NewBH}), scalar field (\ref{phiNew}) and coupling function (\ref{fNew}). 
The non null GB-Tensor components in terms of the coordinate $\tilde{r}=r/\mu^{\frac{1}{4}}$ are given by,
\begin{eqnarray}
&&\Theta_{t}{}^{t}\!=\!\frac{1}{9\mu^{\frac{1}{2}}\tilde{r}^{6}}\!\!\left[ - 3(2\tilde{r}^{4} + 3) + \frac{ 2\tilde{r}^{3}(\tilde{r}^{4}-3)}{(\tilde{r}^{4}-1)^{\frac{5}{6}}}\!\!\int^{\tilde{r}}_{s}\!\!\frac{ d\xi }{ ( \xi^{4} - 1)^{^{\!\!\frac{1}{6}}} }  \right]\!,\quad \Theta_{r}{}^{r}\!=\! \frac{2(2\tilde{r}^{4}-3)}{3\mu^{\frac{1}{2}}\tilde{r}^{6}}\!\!\left[ 3 - \frac{\tilde{r}^{3}}{(\tilde{r}^{4} - 1)^{\frac{5}{6}}}\!\!\int^{\tilde{r}}_{s}\!\!\frac{ d\xi }{ ( \xi^{4} - 1)^{^{\!\!\frac{1}{6}}} } \right]\!, \\
&&\Theta_{\theta}{}^{\theta} \!=\! \Theta_{\varphi}{}^{\varphi} \!=\!\frac{2}{ 9\mu^{\frac{1}{2}}\tilde{r}^{6} }\!\!\left[ - 3(\tilde{r}^{4}-12) + \frac{ \tilde{r}^{3}(\tilde{r}^{4}-3)}{(\tilde{r}^{4}-1)^{\frac{5}{6}}}\!\!\int^{\tilde{r}}_{s}\!\!\frac{ d\xi }{ ( \xi^{4} - 1)^{^{\!\!\frac{1}{6}}} } \right]\!.
\end{eqnarray}
By using (\ref{rho_phi}) and (\ref{Pr_phi}),
\begin{equation}\label{Eq_pr}
- 8\pi\rho - \Theta_{t}{}^{t} 
= \frac{3\mu}{r^{6}}, \quad\quad  8\pi P_{r} - \Theta_{r}{}^{r} 
= \frac{3\mu}{r^{6}}, \quad\quad  8\pi P_{\theta} - \Theta_{\theta}{}^{\theta} 
= -\frac{6\mu}{r^{6}},\quad\quad  8\pi P_{\varphi} - \Theta_{\varphi}{}^{\varphi} 
= -\frac{6\mu}{r^{6}}.
\end{equation}
For the metric the non null Einstein tensor components are; $G_{t}{}^{t}=G_{t}{}^{t}=\frac{3\mu}{r^{6}}$, $G_{\theta}{}^{\theta}=G_{\varphi}{}^{\varphi}=-\frac{6\mu}{r^{6}}$. Therefore, according to (\ref{diagonalEab}), given that, $\rho=-E_{t}{}^{t},$ $P_{r}=E_{r}{}^{r}$, $P_{\theta}=E_{\theta}{}^{\theta}$, $P_{\varphi}=E_{\varphi}{}^{\varphi}$, from (\ref{Eq_pr}) it follows that the field equations $G_{\alpha}{}^{\beta} + \Theta_{\alpha}{}^{\beta} = 8\pi E_{\alpha}{}^{\beta}$ are satisfied.

{\bf $\phi$ and $\phi'$ as function of $\tilde{r}$:}
\begin{eqnarray}
\phi &=& \int^{\tilde{r}}_{\tilde{\sigma}}\left[ \frac{2}{9\chi^{2}(\chi^{4} - 1)}\left(  42\chi^{4} - 45  - \frac{2\chi^{3}(7\chi^{4} - 12)}{ (\chi^{4} - 1)^{\frac{5}{6}} }\!\!\int^{\chi}_{s}\!\!\frac{ d\xi }{ ( \xi^{4} - 1)^{^{\!\!\frac{1}{6}}} } \right) \right]^{\frac{1}{2}}d\chi, \quad\quad \tilde{\sigma}=\sigma/\mu^{\frac{1}{4}}\!,\\
\phi'^{2} &=& \frac{2}{9\mu^{\frac{1}{2}} \tilde{r}^{2} ( \tilde{r}^{4} - 1 )}\left[ 42\tilde{r}^{4} - 45 - \frac{ 2\tilde{r}^{3}(7\tilde{r}^{4} - 12)}{ (\tilde{r}^{4} - 1)^{\frac{5}{6}} }\!\!\int^{\tilde{r}}_{s}\!\!\frac{ d\xi }{ ( \xi^{4} - 1)^{^{\!\!\frac{1}{6}}} }   \right]\!.
\end{eqnarray}
The behavior of $\phi'^{2}$ for different values of $s$, is illustrated in Fig. (\ref{D1phi}).

\end{document}